\documentstyle[twocolumn,aps,pra,array,epsf]{revtex}
\begin{document}
\draft
\title{Multiparty key distribution and secret sharing based on
entanglement swapping}
\author{Ad\'{a}n Cabello\thanks{Electronic address:
adan@cica.es, fite1z1@sis.ucm.es}}
\address{Departamento de F\'{\i}sica Aplicada II,
Universidad de Sevilla, 41012 Sevilla, Spain}
\date{\today}
\maketitle
\begin{abstract}
A general proof
of the security against eavesdropping of a previously
introduced protocol for
two-party quantum key distribution based on entanglement swapping
[Phys. Rev. A {\bf 61}, 052312 (2000)] is provided.
In addition, the protocol is extended to permit multiparty quantum key
distribution and secret sharing of classical information.
\end{abstract}
\pacs{PACS numbers: 03.67.Dd, 03.67.Hk, 03.65.Bz}

\narrowtext
\section{Introduction}
\label{sec:I}
Entanglement swapping (ES), that is,
entangling a set of particles $S$ by
appropriately projecting other particles previously entangled with
particles of $S$ \cite{es,es2,BVK98,es3,BEZ00},
has found a number of
applications in quantum information:
constructing a quantum telephone exchange,
speeding up the distribution of entanglement, correcting errors in
Bell states, and preparing entangled
states of a higher number of particles \cite{BVK98,BEZ00}.
Recently, ES has also been used to solve
the problem of cryptographic key distribution between two parties
in an essentially new way \cite{Cabello00b}.
In this paper, the scheme of Ref.~\cite{Cabello00b} is
extended to permit the distribution of the same key
to several users (multiparty key distribution),
and to permit the distribution
of the same key to several users
in such a way that they must cooperate
to obtain the key
(secret sharing of classical information).
The structure of the paper is the following:
In Sec.~\ref{subsec:IIA} the protocol for quantum key distribution
between two parties based on ES is reviewed.
A general proof of its security is provided in Sec.~\ref{subsec:IIB}.
Multiparty key distribution is introduced,
and a new protocol using
Greenberger-Horne-Zeilinger (GHZ) states \cite{GHZ89}
is presented in Sec.~\ref{subsec:IIIA}.
In Sec.~\ref{subsec:IIIB} a different approach, based on ES and
which is a generalization
the two-party protocol, is introduced.
Secret sharing is treated in Sec.~\ref{sec:IV}.
Two previous protocols for secret sharing
using, respectively,
GHZ and Bell states are
reviewed in Secs.~\ref{subsec:IVA} and \ref{subsec:IVB}. In
Sec.~\ref{subsec:IVC}, it is shown how the scheme of
three-party key distribution based on ES
described in Sec.~\ref{subsec:IIIB},
also permits secret sharing.
The case of more than three users
is treated in Sec.~\ref{sec:V}.
Sec.~\ref{sec:VI} is dedicated to demonstrate the
security of ES-based protocols against eavesdropping.
Finally, the main advantages of these protocols
are summarized in Sec.~\ref{sec:VII}.

\section{Quantum key distribution between two parties}
\label{sec:II}
\subsection{The protocol based on entanglement swapping}
\label{subsec:IIA}
The key distribution problem of cryptography is the following:
Alice wishes to convey a sequence of random classical bits (a ``key'')
to Bob, while preventing that
Eve acquires information without being detected.
This problem, which has no solution by classical means,
can be solved using quantum mechanics \cite{BB84}.
Indeed, subsequent developments have shown that
quantum mechanics provides
different tools to solve the problem.
Some are based on the impossibility of cloning
unknown nonorthogonal quantum states \cite{BB84,B92},
some also use entanglement between two particles
\cite{E91,BBM92},
some combine quantum techniques with classical private amplification
and compression techniques \cite{CL98}, and some are based on
splitting the information in several qubits to which Eve has only
a sequential access \cite{GV95,KI97,Cabello00d}.
In Ref.~\cite{Cabello00b}, a new method
for key distribution based on ES
was introduced.
Let us start by briefly reviewing how this ES-based
protocol works.
Consider the orthonormal basis of Bell states given by:
\begin{eqnarray}
\left| 00 \right\rangle_{ij} & = &
{1 \over \sqrt{2}}
\left( {
\left| 0 \right\rangle_{i} \otimes
\left| 0 \right\rangle_{j} +
\left| 1 \right\rangle_{i} \otimes
\left| 1 \right\rangle_{j}
} \right),
\label{Bell1} \\
\left| 01 \right\rangle_{ij} & = &
{1 \over \sqrt{2}}
\left( {
\left| 0 \right\rangle_{i} \otimes
\left| 0 \right\rangle_{j} -
\left| 1 \right\rangle_{i} \otimes
\left| 1 \right\rangle_{j}
} \right),
\label{Bell2} \\
\left| 10 \right\rangle_{ij} & = &
{1 \over \sqrt{2}}
\left( {
\left| 0 \right\rangle_{i} \otimes
\left| 1 \right\rangle_{j} +
\left| 1 \right\rangle_{i} \otimes
\left| 0 \right\rangle_{j}
} \right),
\label{Bell3} \\
\left| 11 \right\rangle_{ij} & = &
{1 \over \sqrt{2}}
\left( {
\left| 0 \right\rangle_{i} \otimes
\left| 1 \right\rangle_{j} -
\left| 1 \right\rangle_{i} \otimes
\left| 0 \right\rangle_{j}
} \right),
\label{Bell4}
\end{eqnarray}
where
\begin{eqnarray}
{\sigma_z} \left| 0 \right\rangle & = &
\left| 0 \right\rangle,
\label{z1} \\
{\sigma_z} \left| 1 \right\rangle & = &
-\left| 1 \right\rangle,
\label{z2}
\end{eqnarray}
being $\sigma_z$ the corresponding Pauli spin matrix.
For convenience, we shall divide the protocol in three parts:

(I) {\em Preparation}.
Initially, Alice has four qubits: qubits 1 and 2,
prepared in one public Bell state of the basis
(\ref{Bell1})-(\ref{Bell4}), and qubits 3 and 5, prepared in
another Bell public state of the same basis.
Bob, in a distant place, has
two qubits, 4 and 6, prepared in a public Bell state
of the same basis.
For example, the initial state of the six qubits can be
\begin{equation}
\left| \Psi_I \right\rangle =
\left| 00 \right\rangle_{12} \otimes
\left| 00 \right\rangle_{35} \otimes
\left| 00 \right\rangle_{46}.
\label{inicialcero}
\end{equation}
Next, Alice sends qubit 2 out
to Bob through an insecure quantum channel
(i.e., Eve can manipulate qubit 2).

(II) {\em Generation of two bits of the key}.
Alice performs a complete Bell-state
measurement on qubits 1 and 3
(henceforth referred to as Alice's {\em secret}
measurement). The result, $AS$ (a random number: ``00'' if
the result is $\left| 00
\right\rangle$, ``01'' if it is $\left| 01 \right\rangle$, ``10'' if it
is $\left| 10 \right\rangle$, or ``11'' if it is $\left| 11 \right\rangle$),
defines two bits of the key. Then, the state is
\begin{equation}
\left| \Psi_{II} \right\rangle =
\left| AS \right\rangle_{13} \otimes
\left| AS' \right\rangle_{25} \otimes
\left| 00 \right\rangle_{46},
\end{equation}
where $\left| AS' \right\rangle$ is a Bell state which
is in one-to-one correspondence
with $\left| AS\right\rangle$.

(III) {\em How Bob obtains the two bits of the key}.
Bob performs a complete Bell-state measurement on qubits 2 and 4
(henceforth referred to as Bob's {\em secret} measurement),
and keeps the result, $BS$, secret.
After that, the state is
\begin{equation}
\left| \Psi_{III} \right\rangle =
\left| AS \right\rangle_{13} \otimes
\left| BS \right\rangle_{24} \otimes
\left| AP \right\rangle_{56},
\end{equation}
where $\left| AP \right\rangle$ is a Bell state which can be determined from
the pair $AS$, $BS$.
Then, Bob sends qubit 6 out to
Alice (i.e., Eve can manipulate qubit 6). Finally,
Alice performs a complete Bell-state
measurement on qubits 5 and 6 (henceforth
referred to as Alice's {\em public}
measurement), and publicly announces the
result, $AP$, through a classical channel
(which is assumed to be public but which
cannot be altered) \cite{com}.
Due to the successive ES between the pairs of
qubits, for each of the four possible values of $AP$,
there is a different one-to-one correspondence between the results
of Alice's and Bob's
secret measurements.
These correspondences are compiled in Table I.
Therefore, once Bob knows the
$AP$, he can infer $AS$.
The process must be sequentially repeated until
the key is large enough.

\subsection{Security of the protocol based on ES}
\label{subsec:IIB}
In Ref.~\cite{Cabello00b}, the security
against eavesdropping of the
protocol for two-party key distribution based on
ES was showed for a particular
eavesdropping attack.
Here I will provide a general (i.e., attack-independent) proof.

The result $AS$ defines two bits of the key.
However, $AS$ is random
and Eve cannot influence it by manipulating any of
the transmitted qubits.
To obtain $AS$, Eve needs the same two ingredients as Bob:
$BS$ and $AP$. In addition, to avoid being detected,
Eve needs to obtain $BS$ {\em without changing} $AP$.
However, since Eve has only access to
two of the six qubits, we shall see that any procedure that
allows Eve to obtain $BS$, changes $AP$ in an unpredictable way.
Let us examine the strategies that Eve can follow and their
consequences.

In step (I) of the protocol,
the only qubit accessible to Eve is qubit 2.
If Eve's aim is to obtain $BS$ (as a previous step to obtain $AS$),
the only useful strategy is one whose result is equivalent to
capturing qubit 2 and substituting it by an ancillary qubit 8
(which will be sent out to Bob), which was previously prepared
in a Bell state (for instance $\left| 00 \right\rangle_{78}$)
with another ancillary qubit 7
(which will be retained by Eve) \cite{com2}.
After this man{\oe}uvre the state of the qubits is
\begin{equation}
\left| \Psi'_I \right\rangle =
\left| 00 \right\rangle_{12} \otimes
\left| 00 \right\rangle_{35} \otimes
\left| 00 \right\rangle_{46} \otimes
\left| 00 \right\rangle_{78},
\end{equation}
where Alice has qubits 1, 3, and 5; Bob has qubits 4, 6, and 8; and Eve
has qubits 2 and 7. This situation is illustrated in Fig.~1 (a1).
The corresponding situation in the alternative scenario in which
Eve is not present is illustrated in Fig.~1 (b1).

In step (II), Alice performs her secret measurement on qubits 1 and 3,
and Bob performs his secret measurement on qubits 4 and 8
(which substitutes qubit 2).
After these measurements the state of the qubits is
\begin{equation}
\left| \Psi'_{II} \right\rangle =
\left| AS \right\rangle_{13} \otimes
\left| AS' \right\rangle_{25} \otimes
\left| BS' \right\rangle_{67} \otimes
\left| BS \right\rangle_{48},
\end{equation}
where $\left| AS \right\rangle$ is the Bell state which defines two bits
of the key, $\left| AS' \right\rangle$ is a Bell state in one-to-one
correspondence with $\left| AS \right\rangle$, $\left| BS \right\rangle$
is the Bell state which gives Bob's secret result $BS$,
and $\left| BS' \right\rangle$ is a Bell state in one-to-one correspondence
with $\left| BS \right\rangle$. This situation is illustrated in Fig.~1 (a2).
The corresponding situation
in the alternative scenario in which Eve is not
present is illustrated in Fig.~1 (b2).

In the step (III), Bob sends qubit out 6 to Alice.
This allows Eve to capture it and obtain $BS$ by performing a Bell-state
measurement on qubits 6 and 7 (the result $BS'$ of this measurement is
in one-to-one correspondence with $BS$). This is (modulo equivalencies)
the only strategy that allows Eve to obtain $BS$. Now, to obtain
$AS$, she still needs to know $AP$
(which must be in one-to-one correspondence with the pair $AS$, $BS$).
However, Eve's intervention has changed the state of the qubits
(compare $\left| \Psi_{II} \right\rangle$ with
$\left| \Psi'_{II} \right\rangle$).

Before Alice's public measurement, Eve has access to qubit 2
(which is in a Bell state, unknown to Eve, with Eve's qubit 5), and to
qubits 6 and 7. If Eve manages to give Alice a qubit in the Bell state
$\left| AP \right\rangle$ with qubit 2,
then her intervention won't be detected.
Eve can prepare a qubit in any desired Bell state with
qubit 2; the problem is that she does not know which is the ``correct''
Bell state. Indeed, she cannot know which is the right one
since this would require Eve to know $\left| AS \right\rangle_{13}$
(and Eve has no access to qubits 1 and 3), or
$\left| AS' \right\rangle_{25}$ (and Eve only has access to qubit 2,
and since the partial trace of all the Bell states is the identity matrix,
any measurement on one qubit cannot reveal anything about the state
of both qubits).

Alternatively, if Eve gives to Alice qubit 6 (or 7),
then the result of Alice's
public measurement will allow
Eve to obtain $AS'$ (and therefore $AS$).
However, this result is not
in one-to-one correspondence with the pair $AS$, $BS$ anymore.
Therefore, the result obtained by Alice will
be the ``wrong'' one in $\frac{3}{4}$ of the runs,
and thus Eve's intervention can be detected when
Alice and Bob compare subsets of their keys.

Summing up, no strategy allows Eve to extract information
without being detected, because the only strategy that Eve can use
to obtain information will change the expected result for $AP$
in $\frac{3}{4}$ of the cases. In addition, this proves that one
of the interesting features of the protocol based on ES
---namely, that it improves the efficiency of the
detection of eavesdropping compared with other
protocols \cite{Cabello00b}--- is
independent of the attack.

\section{Multiparty key distribution}
\label{sec:III}
\subsection{Multiparty key distribution using two GHZ states}
\label{subsec:IIIA}
Consider the following problem:
Alice wishes to convey the {\em same} key
to $N$ users (Bob, Carol,\dots, Nathan), while preventing
Eve from acquiring information without being detected.
This problem, called {\em multiparty key distribution},
is a special case of networked
{\em cryptographic conferencing} \cite{conf,conf2}.

Here I introduce a protocol for using GHZ states
for multiparty quantum key distribution that,
as far as I know, has not been presented anywhere before.
It can be considered as a generalization to many parties
of the two-party protocol of Ref.~\cite{E91}.

Let us focus our attention in the case $N=3$
(the cases with $N>3$ are straightforward extensions of this case).
Alice wishes to
distribute the same key to Bob and Carol.
For that purpose,
she randomly prepares one of the following two three-qubit
GHZ states:
\begin{eqnarray}
\left| {\psi_z} \right\rangle_{ijk} & = &
{1 \over \sqrt{2}}
(
\left| 0 \right\rangle_{i} \otimes
\left| 0 \right\rangle_{j} \otimes
\left| 0 \right\rangle_{k} + \nonumber \\
& & \left| 1 \right\rangle_{i} \otimes
\left| 1 \right\rangle_{j} \otimes
\left| 1 \right\rangle_{k}
),
\label{GHZz} \\
\left| {\psi_x} \right\rangle_{ijk} & = &
{1 \over \sqrt{2}}
(
\left| \bar{0} \right\rangle_{i} \otimes
\left| \bar{0} \right\rangle_{j} \otimes
\left| \bar{0} \right\rangle_{k} + \nonumber \\
& & \left| \bar{1} \right\rangle_{i} \otimes
\left| \bar{1} \right\rangle_{j} \otimes
\left| \bar{1} \right\rangle_{k}
),
\end{eqnarray}
where
\begin{eqnarray}
{\sigma_x} \left| \bar{0} \right\rangle & = &
\left| \bar{0} \right\rangle,
\label{x1} \\
{\sigma_x} \left| \bar{1} \right\rangle & = &
-\left| \bar{1} \right\rangle,
\label{x2}
\end{eqnarray}
being ${\sigma_x}$ the corresponding Pauli spin matrix.
Then Alice sends one of the three qubits out to Bob,
another to Carol,
and retains the third one.
Bob and Carol perform a measurement of
either ${\sigma_z}$ or ${\sigma_x}$
on their own qubit. When Alice has prepared the state
$\left| \psi_z \right\rangle$ ($\left| \psi_x \right\rangle$),
and both Bob and Carol have measured
$\sigma_z$ ($\sigma_x$)
---i.e., in $1\over{4}$ of the cases---,
all of them obtain the same result.
In that case, they can use this result
to define one bit of the key.
The other cases are not useful for establishing a common key
and are rejected.
Alternatively, to reduce the wastage of qubits due to the
noncoincidence of the measurements, Alice can tell
Bob and Carol which is the ``right'' measurement
(once all three
qubits are safe from Eve's intervention).
The security of this scheme against
eavesdropping is guaranteed by
the impossibility of cloning an unknown state chosen between
$\left| \psi_z \right\rangle$ and $\left| \psi_x \right\rangle$,
specially when Eve only has
access to two of the three qubits.

\subsection{Multiparty key distribution
based on entanglement swapping}
\label{subsec:IIIB}
A different multiparty key distribution protocol
can be obtained using ES.
Indeed, what follows is just one of
the possible generalizations to three parties of the protocol for
key distribution between two parties
based on ES of Ref.~\cite{Cabello00b}.

Consider the orthonormal basis of GHZ states given by:
\begin{eqnarray}
\left| 000 \right\rangle_{ijk} & = &
{1 \over \sqrt{2}}
(
\left| 0 \right\rangle_{i} \otimes
\left| 0 \right\rangle_{j} \otimes
\left| 0 \right\rangle_{k} + \nonumber \\
 & & \left| 1 \right\rangle_{i} \otimes
\left| 1 \right\rangle_{j} \otimes
\left| 1 \right\rangle_{k}
),
\label{GHZ1} \\
\left| 001 \right\rangle_{ijk} & = &
{1 \over \sqrt{2}}
(
\left| 0 \right\rangle_{i} \otimes
\left| 0 \right\rangle_{j} \otimes
\left| 0 \right\rangle_{k} - \nonumber \\
 & & \left| 1 \right\rangle_{i} \otimes
\left| 1 \right\rangle_{j} \otimes
\left| 1 \right\rangle_{k}
), \\
\left| 010 \right\rangle_{ijk} & = &
{1 \over \sqrt{2}}
(
\left| 0 \right\rangle_{i} \otimes
\left| 0 \right\rangle_{j} \otimes
\left| 1 \right\rangle_{k} + \nonumber \\
 & & \left| 1 \right\rangle_{i} \otimes
\left| 1 \right\rangle_{j} \otimes
\left| 0 \right\rangle_{k}
), \\
\left| 011 \right\rangle_{ijk} & = &
{1 \over \sqrt{2}}
(
\left| 0 \right\rangle_{i} \otimes
\left| 0 \right\rangle_{j} \otimes
\left| 1 \right\rangle_{k} - \nonumber \\
 & & \left| 1 \right\rangle_{i} \otimes
\left| 1 \right\rangle_{j} \otimes
\left| 0 \right\rangle_{k}
), \\
\left| 100 \right\rangle_{ijk} & = &
{1 \over \sqrt{2}}
(
\left| 0 \right\rangle_{i} \otimes
\left| 1 \right\rangle_{j} \otimes
\left| 0 \right\rangle_{k} + \nonumber \\
 & & \left| 1 \right\rangle_{i} \otimes
\left| 0 \right\rangle_{j} \otimes
\left| 1 \right\rangle_{k}
), \\
\left| 101 \right\rangle_{ijk} & = &
{1 \over \sqrt{2}}
(
\left| 0 \right\rangle_{i} \otimes
\left| 1 \right\rangle_{j} \otimes
\left| 0 \right\rangle_{k} - \nonumber \\
 & & \left| 1 \right\rangle_{i} \otimes
\left| 0 \right\rangle_{j} \otimes
\left| 1 \right\rangle_{k}
), \\
\left| 110 \right\rangle_{ijk} & = &
{1 \over \sqrt{2}}
(
\left| 1 \right\rangle_{i} \otimes
\left| 0 \right\rangle_{j} \otimes
\left| 0 \right\rangle_{k} + \nonumber \\
 & & \left| 0 \right\rangle_{i} \otimes
\left| 1 \right\rangle_{j} \otimes
\left| 1 \right\rangle_{k}
), \\
\left| 111 \right\rangle_{ijk} & = &
{1 \over \sqrt{2}}
(
\left| 1 \right\rangle_{i} \otimes
\left| 0 \right\rangle_{j} \otimes
\left| 0 \right\rangle_{k} - \nonumber \\
 & & \left| 0 \right\rangle_{i} \otimes
\left| 1 \right\rangle_{j} \otimes
\left| 1 \right\rangle_{k}
).
\label{GHZ8}
\end{eqnarray}
The protocol can be summarized in four steps,
which are illustrated in Fig.~2:

(i) Initially, Alice has qubits 1 and 2,
prepared in one public Bell state of the basis
(\ref{Bell1})-(\ref{Bell4}), and qubits 3, $A$, and $B$
(qubits described by numbers stay with the same user
during all the protocol, and qubits described by letters
are transmitted between users during the protocol)
prepared in a GHZ state of the basis (\ref{GHZ1})-(\ref{GHZ8}).
Bob (Carol), in a distant place,
has two qubits, 5 and $D$ (4 and $C$),
prepared in a public Bell state.
For instance, the initial state of the nine qubits can be
\begin{equation}
\left| {\psi_i} \right\rangle =
\left| 000 \right\rangle_{3AB} \otimes
\left| 00 \right\rangle_{12} \otimes
\left| 00 \right\rangle_{5D} \otimes
\left| 00 \right\rangle_{4C},
\label{inicial}
\end{equation}
where subindexes 3, $A$, etc., mean qubits 3, $A$, etc.

(ii) Then,
Alice sends qubit $A$ ($B$) out to Bob (Carol)
through an insecure quantum channel.
Next, Alice performs a secret Bell-state
measurement on qubits 2 and 3,
Bob performs a secret Bell-state measurement
on qubits 5 and $A$, and Carol
performs a secret Bell-state measurement on qubits
4 and $B$.

(iii) After these three secret measurements,
the state of qubits 1, $C$, and $D$
becomes a GHZ state of the basis
(\ref{GHZ1})-(\ref{GHZ8}), due to multiparticle ES
\cite{BVK98}. The final state is
\begin{equation}
\left| {\psi_{iii}} \right\rangle =
\left| AP \right\rangle_{1CD} \otimes
\left| AS \right\rangle_{23} \otimes
\left| BS \right\rangle_{5A} \otimes
\left| CS \right\rangle_{4B}.
\end{equation}

(iv) Then, Bob (Carol) sends qubit $D$ ($C$) out to Alice,
who performs a complete GHZ-state measurement
on qubits 1, $C$, and $D$ [i.e., a
measurement which unambiguously discriminates between states
(\ref{GHZ1})-(\ref{GHZ8})], and publicly
announces the result through a
classical channel.

Out of the 512 possible combinations of results
(for Alice's public measurement, and Alice's, Bob's, and Carol's
secret measurements),
there are only 64
which have a nonzero probability to occur.
If the initial state is (\ref{inicial}), these 64 combinations
are represented in Table II. All
of them have the same probability to occur ($\frac{1}{64}$).

The secret key that Alice, Bob, and Carol will share is
defined as the {\em first} bit of Alice's secret measurement.
As a close examination of Table II reveals,
Bob (or Carol) can infer Alice's first bit
using just two ingredients: the result of the public measurement, and
the result of his (her) own secret measurement.
Therefore, once Bob (Carol) knows
the result of the public measurement, he (she) can infer the
first bit of the result of Alice's
secret measurement.
The process can be sequentially repeated.

\section{Quantum secret sharing of classical information}
\label{sec:IV}
\subsection{Hillery-Bu\v{z}ek-Berthiaume
secret sharing using GHZ states}
\label{subsec:IVA}
Consider the following problem:
Alice wishes to convey a key
to Bob and Carol in such a way that none of them can read
it on their own,
only if they collaborate.
In addition, they wish to prevent that
Eve acquires information without being detected.
This is an interesting problem in the following scenario \cite{HBB99}:
Alice wants to have a secret action taken on her behalf in a distant part.
There she has two agents, Bob and Carol, who carry it out for her.
Alice knows that one and only one of them is dishonest, but she does
not know which one.
She cannot simply send a secure message to both of them,
because the dishonest one will try to sabotage the action,
but she knows that if both carry it out together,
the honest one will keep the dishonest one from doing
any damage.

A first solution to this problem using quantum tools was provided in
Ref.~\cite{HBB99}, and can be summarized as follows:
Alice prepares three qubits in
the GHZ state given by Eq.~(\ref{GHZz}), and sends one qubit out
to Bob, another to Carol, and keeps the third.
Bob and Carol independently and randomly choose whether
to measure $\sigma_x$ or $\sigma_y$
on their qubits.
They then publicly announce which measurement they have made,
but not which result they have obtained.
If Bob and Carol have chosen the same measurement,
they can then determine what was
the result of Alice's measurement by
combining their results.
This allows Alice, Bob, and Carol to establish a common key.
The other events in which Bob and Carol have chosen
different measurements (which are $\frac{1}{2}$ of the events)
do not allow them to make useful inferences to establish a key
and are therefore rejected.
For details on this protocol and for proofs of its security
see \cite{HBB99,KKI99}.

\subsection{Karlsson-Koashi-Imoto secret sharing using Bell states}
\label{subsec:IVB}
In Ref.~\cite{KKI99} another protocol
for secret sharing using Bell states
instead of GHZ states is proposed.
It works as follows:
Alice prepares two qubits in one of the following four states:
\begin{eqnarray}
\left| {\psi^+} \right\rangle_{ij} & = &
{1 \over \sqrt{2}}
\left( {
\left| 0 \right\rangle_{i} \otimes
 \left| 0 \right\rangle_{j} +
\left| 1 \right\rangle_{i} \otimes
\left| 1 \right\rangle_{j}
} \right)
\nonumber \\
& = &
{1 \over \sqrt{2}}
\left( {
\left| \bar{0} \right\rangle_{i} \otimes
\left| \bar{0} \right\rangle_{j} -
\left| \bar{1} \right\rangle_{i} \otimes
\left| \bar{1} \right\rangle_{j}
} \right),
\label{Be1} \\
\left| {\phi^-} \right\rangle_{ij} & = &
{1 \over \sqrt{2}}
\left( {
\left| 0 \right\rangle_{i} \otimes
\left| 0 \right\rangle_{j} -
\left| 1 \right\rangle_{i} \otimes
\left| 1 \right\rangle_{j}
} \right)
\nonumber \\ & = &
{1 \over \sqrt{2}}
\left( {
\left| \bar{0} \right\rangle_{i} \otimes
\left| \bar{1} \right\rangle_{j}+
\left| \bar{1} \right\rangle_{i} \otimes
\left| \bar{0} \right\rangle_{j}
} \right),
\label{Be2} \\
\left| {\Psi^+} \right\rangle_{ij} & = &
{1 \over \sqrt{2}}
\left( {
\left| 0 \right\rangle_{i} \otimes
\left| \bar{0} \right\rangle_{j} +
\left| 1 \right\rangle_{i} \otimes
\left| \bar{1} \right\rangle_{j}
} \right)
\nonumber \\ & = &
{1 \over \sqrt{2}}
\left( {
\left| \bar{0} \right\rangle_{i} \otimes
\left| 0 \right\rangle_{j} +
\left| \bar{1} \right\rangle_{i} \otimes
\left| 1 \right\rangle_{j}
} \right),
\label{Be3} \\
\left| {\Phi^-} \right\rangle_{ij} & = &
{1 \over \sqrt{2}}
\left( {
\left| 0 \right\rangle_{i} \otimes
\left| \bar{1} \right\rangle_{i} -
\left| 1\right\rangle_{i} \otimes
\left| \bar{0} \right\rangle_{j}
} \right)
\nonumber \\ & = &
{1 \over \sqrt{2}}
\left( {
\left| \bar{0} \right\rangle_{i} \otimes
\left| 1 \right\rangle_{j} -
\left| \bar{1} \right\rangle_{i} \otimes
\left| 0 \right\rangle_{j}
} \right),
\label{Be4}
\end{eqnarray}
and sends out one of the qubits to Bob and the other to Carol.
They independently and randomly perform a measurement of
either $\sigma_z$ or $\sigma_x$.
Then Bob and Carol have a public discussion where
they declare the measurement outcomes for a subset of bits used
for testing eavesdropping.
It is essential that this discussion takes place
before any further declaration.
If they do not detect eavesdropping, Bob publicly declares the
outcomes of his measurements
(but not yet his choice of measurements),
then Carol declares both her choice of measurements
and the corresponding outcomes,
and finally Bob declares his choice of measurements (the order
of the declarations is important to preserve security).
Then, Alice publicly reveals
whether she has prepared one of the states
$\left \{ \left| \psi^+ \right\rangle,\,\left| \phi^- \right\rangle \right\}$,
or one of the states
$\left \{ \left| \Psi^+ \right\rangle,\,\left| \Phi^- \right\rangle \right\}$
(but not which specific state she has prepared).
If Alice has prepared a state of the first (second) set, then
the results of Bob's and Carol's
local measurements are correlated only if
both have chosen to measure $\sigma_z$ or both
have chosen to measure $\sigma_x$
(if one of them has chosen to measure $\sigma_z$
and the other has chosen to measure $\sigma_x$).
Such correlations allow Bob and Carol to find out
which state Alice has prepared. But this is only possible if
both cooperate.
Note that in this protocol,
in half of the events there is no correlation between
Bob's and Carol's results so half of the events are
not useful for secret sharing and must be rejected.

\subsection{Secret sharing using entanglement swapping}
\label{subsec:IVC}
In Sec.~\ref{subsec:IIIB} we saw how
to distribute one bit between three users
employing ES between two-qubit Bell states and three-qubit
GHZ states. In this section we show that
the scenario described there also allows
secret sharing of classical information.
The protocol for secret sharing has
steps (i) to (iii) in common with the protocol
of multiparty key distribution described
in Sec.~\ref{subsec:IIIB}. In step (iv), we saw that
once Bob (Carol) knows
the result of the public measurement, he (she) can infer the
{\em first} bit of the result of Alice's
secret measurement. In addition, as a close inspection of
Table II shows, once Bob (Carol) knows the result of Carol's
(Bob's) secret measurement, he (she) can infer the
{\em second} bit of the result of Alice's secret measurement.
That is, if Bob and Carol cooperate they can infer this second bit.
Therefore, the same scenario allows us to
develop a protocol for multiparty
key distribution and, simultaneously,
a protocol for secret sharing.

\section{ES-based protocol for multiparty key distribution and
secret sharing between more than three users}
\label{sec:V}
Both the scheme for multiparty key distribution based on ES,
described in Sec.~\ref{subsec:IIIB}, and the scheme for secret sharing
described in Sec.~~\ref{subsec:IVC}, can be extended to $N$ users
as follows:

(i) Every user has a pair of qubits in a public
Bell state. In addition, Alice has another $N$ qubits prepared
in a GHZ state of the orthonormal basis:
\begin{eqnarray}
\left| 00...0 \right\rangle_{ij...N} & = &
{1 \over \sqrt{2}}
(
\left| 0 \right\rangle_{i} \otimes
\left| 0 \right\rangle_{j} \otimes ... \otimes
\left| 0 \right\rangle_{N} + \nonumber \\
 & & \left| 1 \right\rangle_{i} \otimes
\left| 1 \right\rangle_{j} \otimes ... \otimes
\left| 1 \right\rangle_{N}
),
\label{GHZN1} \\
\left| 00...1 \right\rangle_{ij...N} & = &
{1 \over \sqrt{2}}
(
\left| 0 \right\rangle_{i} \otimes
\left| 0 \right\rangle_{j} \otimes ... \otimes
\left| 0 \right\rangle_{N} - \nonumber \\
 & & \left| 1 \right\rangle_{i} \otimes
\left| 1 \right\rangle_{j} \otimes ... \otimes
\left| 1 \right\rangle_{N}
), \\
 & ... & \nonumber \\
\left| 11...0 \right\rangle_{ij...N} & = &
{1 \over \sqrt{2}}
(
\left| 1 \right\rangle_{i} \otimes
\left| 0 \right\rangle_{j} \otimes ... \otimes
\left| 0 \right\rangle_{N} + \nonumber \\
 & & \left| 0 \right\rangle_{i} \otimes
\left| 1 \right\rangle_{j} \otimes ... \otimes
\left| 1 \right\rangle_{N}
), \\
\left| 11...1 \right\rangle_{ij...N} & = &
{1 \over \sqrt{2}}
(
\left| 1 \right\rangle_{i} \otimes
\left| 0 \right\rangle_{j} \otimes ... \otimes
\left| 0 \right\rangle_{N} - \nonumber \\
 & & \left| 0 \right\rangle_{i} \otimes
\left| 1 \right\rangle_{j} \otimes ... \otimes
\left| 1 \right\rangle_{N}
).
\label{GHZN8}
\end{eqnarray}
For instance, consider that the initial state of the system is
\begin{equation}
\left| \Psi_i \right\rangle =
\left| 00...0 \right\rangle \otimes
\left| 00 \right\rangle \otimes ... \otimes
\left| 00 \right\rangle.
\end{equation}

(ii) Then, Alice sends a qubit of her GHZ state out to each of the other
$N-1$ users. Next, each user (including Alice) performs a Bell-state
measurement on the received qubit and one of their qubits.

(iii) After these measurements the state of the system becomes
\begin{equation}
\left| \Psi_{iii} \right\rangle =
\left| AP \right\rangle \otimes
\left| AS \right\rangle \otimes
\left| BS \right\rangle \otimes ... \otimes
\left| NS \right\rangle,
\end{equation}
where $\left| AP \right\rangle$ is a $N$-qubit GHZ state of the basis
(\ref{GHZN1})-(\ref{GHZN8}), and
the ordering of the qubits is not the same as in $\left| \Psi_{i} \right\rangle$
(as occurs in Secs.~II A and III B).

(iv) Then, the $N-1$ users sends a
qubit (the one they have no used) to Alice,
and she performs a measurement
to discriminate between the $2^N$ GHZ states
(\ref{GHZN1})-(\ref{GHZN8}),
and publicly announces the result.
This result $AP$, and the result of
their own secret measurement allow each
legitimate user to infer the first bit of
Alice's secret result $AS$.
To find out the second bit of $AS$,
{\em all} users (except Alice) must cooperate.
For instance, in case there are four users
(Alice, Bob, Carol, and David),
to obtain the second bit of $AS$
it is not enough that Bob and Carol
share their secret results.
As Table III shows, they
also need to know David's secret result.

\section{Security of the protocols based on multiparticle ES}
\label{sec:VI}
The proof of the security against eavesdropping of the protocols
for multiparty key distribution and
secret sharing based on multiparticle ES
is parallel to the one developed in Sec.~\ref{subsec:IIB} for the
protocol for two parties key distribution. The guidelines of the proof
are the following:

$AS$ (whose first bit defines the part of the key that the legitimate
users can obtain without cooperating, and whose second bit defines
the part of the key that the users can obtain
if they cooperate) is a random
number, and Eve cannot do anything to change or influence it.

In order to obtain the first (second) bit of $AS$, Eve needs the same
ingredients that any legitimate user needs:
the result of the secret measurement
of one (all) of them, and $AP$.

Any attempt to find out one of the secret results
will change the result $AP$ in
an unpredictable way. Therefore, Eve's presence can be detected.
Indeed, detecting Eve requires the comparison of fewer bits than
in other protocols since the
probability that the result $AP$ is a ``wrong''
one is $\frac{2^N-1}{2^N}$, being $N$ the number of users.

\section{Conclusions}
\label{sec:VII}
The main aim of this paper has been to introduce new protocols
for multiparty key distribution and secret
sharing of classical information.
The main interest of the protocols based on ES is that they provide
a conceptually different way to solve certain
problems of information theory.
Its main advantages are that no transmitted quantum
data are rejected, so they improve the efficiency
of previous protocols; and that the detection of Eve requires
the comparison of fewer bits, since the probability
that Eve alters the result expected by the legitimate users
is higher. On the other hand, since these protocols involve complete
Bell-state and GHZ-state discriminations, they are much more
difficult to perform in practice than previous protocols based on simpler
measurements.

In this paper we have focused our attention in the distribution
of classical information. However, as occurs with previous
proposals, the protocols presented here
can also be used, with little modifications, to distribute
quantum information \cite{BVK98,HBB99}
and for secret sharing of quantum
information \cite{HBB99,KKI99,CGL99,Gottesman00}.

\section*{Acknowledgments}
The author thanks S. Bose for useful discussions on
multiparty key distribution and
the organizers of the Sixth Benasque Center for Physics,
the University of Seville, and
the Junta de Andaluc\'{\i}a for support.

\newpage


\begin{table}[tbp]
\begin{center}
\begin{tabular}{|ccc|ccc|}
\hline
Public & Alice & Bob & Public & Alice & Bob\\ \hline
00 & 00 & 00 & 10 & 00 & 10 \\
'' & 01 & 01 & '' & 01 & 11 \\
'' & 10 & 10 & '' & 10 & 00 \\
'' & 11 & 11 & '' & 11 & 01 \\
01 & 00 & 01 & 11 & 00 & 11 \\
'' & 01 & 00 & '' & 01 & 10 \\
'' & 10 & 11 & '' & 10 & 01 \\
'' & 11 & 10 & '' & 11 & 00 \\
\hline
\end{tabular}
\end{center}
\vspace{0.2cm}
\noindent TABLE I. {\small The 16 possible combinations of results
of Alice's public Bell-state measurement,
and Alice's and Bob's secret Bell-state measurements
on the initial state given by Eq.~(\ref{inicialcero}).}
\end{table}

\begin{table}[tbp]
\begin{center}
\begin{tabular}{|cccc|cccc|}
\hline
Public & Alice & Bob & Carol & Public & Alice & Bob & Carol\\ \hline
000 & 00 & 00 & 00 & 100 & 00 & 00 & 10 \\
'' & '' & 01 & 01 & '' & '' & 01 & 11 \\
'' & 01 & 00 & 01 & '' & 01 & 00 & 11 \\
'' & '' & 01 & 00 & '' & '' & 01 & 10 \\
'' & 10 & 10 & 10 & '' & 10 & 10 & 00 \\
'' & '' & 11 & 11 & '' & '' & 11 & 01 \\
'' & 11 & 10 & 11 & '' & 11 & 10 & 01 \\
'' & '' & 11 & 10 & '' & '' & 11 & 00 \\
001 & 00 & 00 & 01 & 101 & 00 & 00 & 11 \\
'' & '' & 01 & 00 & '' & '' & 01 & 10 \\
'' & 01 & 00 & 00 & '' & 01 & 00 & 10 \\
'' & '' & 01 & 01 & '' & '' & 01 & 11 \\
'' & 10 & 10 & 11 & '' & 10 & 10 & 01 \\
'' & '' & 11 & 10 & '' & '' & 11 & 00 \\
'' & 11 & 10 & 10 & '' & 11 & 10 & 00 \\
'' & '' & 11 & 11 & '' & '' & 11 & 01 \\
010 & 00 & 10 & 00 & 110 & 00 & 10 & 10 \\
'' & '' & 11 & 01 & '' & '' & 11 & 11 \\
'' & 01 & 10 & 01 & '' & 01 & 10 & 11 \\
'' & '' & 11 & 00 & '' & '' & 11 & 10 \\
'' & 10 & 00 & 10 & '' & 10 & 00 & 00 \\
'' & '' & 01 & 11 & '' & '' & 01 & 01 \\
'' & 11 & 00 & 11 & '' & 11 & 00 & 01 \\
'' & '' & 01 & 10 & '' & '' & 01 & 00 \\
011 & 00 & 10 & 01 & 111 & 00 & 10 & 11 \\
'' & '' & 11 & 00 & '' & '' & 11 & 10 \\
'' & 01 & 10 & 00 & '' & 01 & 10 & 10 \\
'' & '' & 11 & 01 & '' & '' & 11 & 11 \\
'' & 10 & 00 & 11 & '' & 10 & 00 & 01 \\
'' & '' & 01 & 10 & '' & '' & 01 & 00 \\
'' & 11 & 00 & 10 & '' & 11 & 00 & 00 \\
'' & '' & 01 & 11 & '' & '' & 01 & 01 \\
\hline
\end{tabular}
\end{center}
\vspace{0.2cm}
\noindent TABLE II. {\small The 64 possible combinations of results
of Alice's public GHZ-state measurement,
Alice's, Bob's, and Carol's secret Bell-state measurements
on the initial state given by Eq.~(\ref{inicial}).}
\end{table}

\begin{table}[tbp]
\begin{center}
\begin{tabular}{|ccccc|}
\hline
Public & Alice & Bob & Carol & David\\ \hline
0000 & 00 & 00 & 00 & 00 \\
'' & '' & 00 & 01 & 01 \\
'' & '' & 01 & 00 & 01 \\
'' & '' & 01 & 01 & 00 \\
'' & 01 & 00 & 00 & 01 \\
'' & '' & 00 & 01 & 00 \\
'' & '' & 01 & 00 & 00 \\
'' & '' & 01 & 01 & 01 \\
'' & 10 & 10 & 10 & 10 \\
'' & '' & 10 & 11 & 11 \\
'' & '' & 11 & 10 & 11 \\
'' & '' & 11 & 11 & 10 \\
'' & 11 & 10 & 10 & 11 \\
'' & '' & 10 & 11 & 10 \\
'' & '' & 11 & 10 & 10 \\
'' & '' & 11 & 11 & 11 \\
\hline
\end{tabular}
\end{center}
\vspace{0.2cm}
\noindent TABLE III. {\small The 16 possible combinations of results
of Alice's, Bob's, Carol's, and David's secret Bell-state measurements
on the initial state
$\left| 0000 \right\rangle \otimes
\left| 00 \right\rangle \otimes
\left| 00 \right\rangle \otimes
\left| 00 \right\rangle \otimes
\left| 00 \right\rangle$, if the result
of Alice's public four-qubit GHZ-state measurement is ``0000''.}
\end{table}

\newpage


\begin{figure}
\epsfxsize=8.6cm
\epsfbox{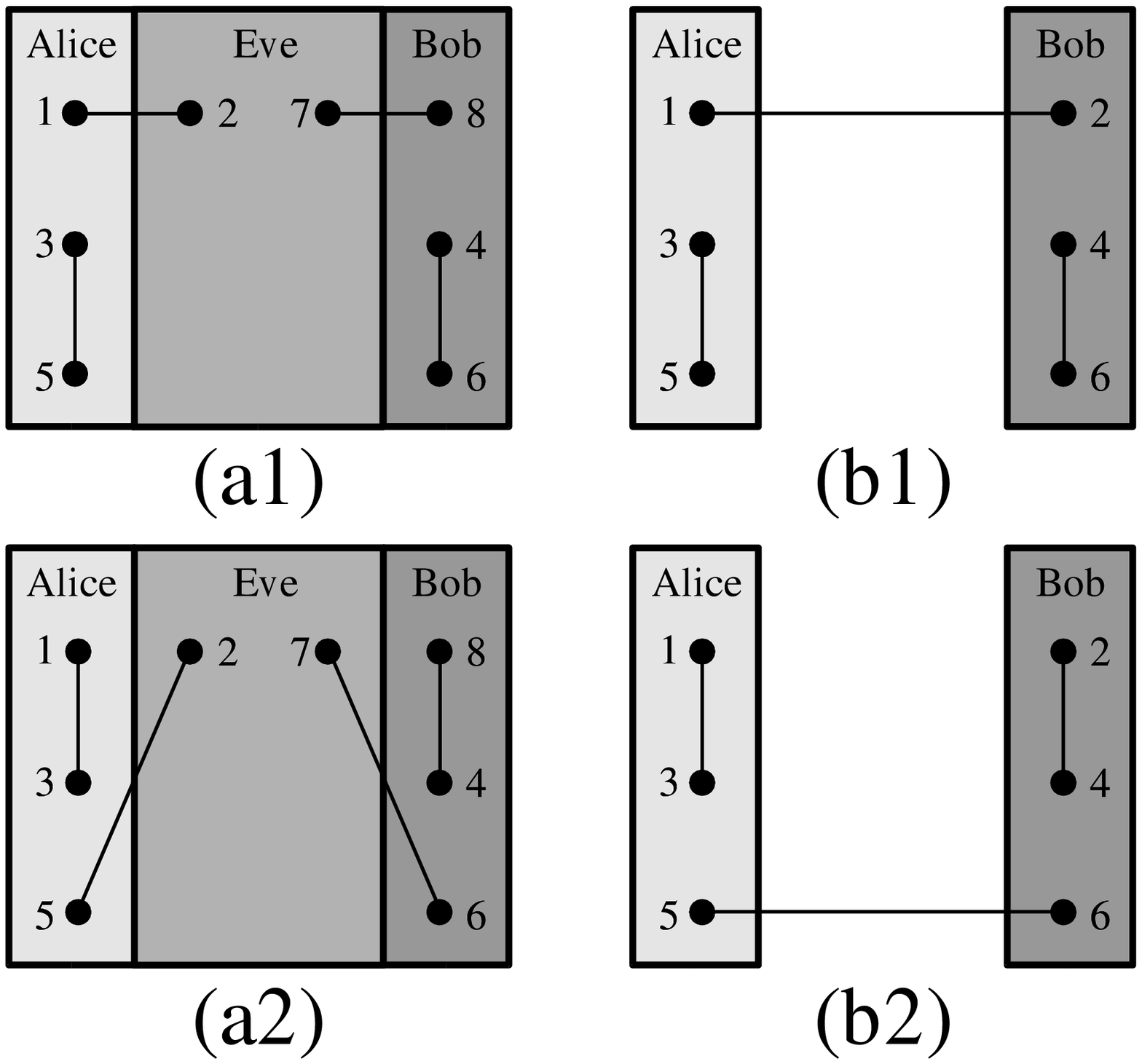}
\end{figure}
\noindent FIG.~1: {\small (a1) and (a2) represent two steps of
the protocol for two-party key distribution
based on ES, assuming that Eve wants to obtain the result of
Bob's secret measurement. (a1) represents the situation before
Alice's and Bob's secret measurements, and (a2) the situation after
these measurements. (b1) and (b2) represent, respectively, the
same two steps, (a1) and (a2), but assuming that Eve is not present.
Bold lines connect qubits in Bell states.}

\begin{figure}
\epsfysize=20.8cm
\hspace{1.2cm}
\epsfbox{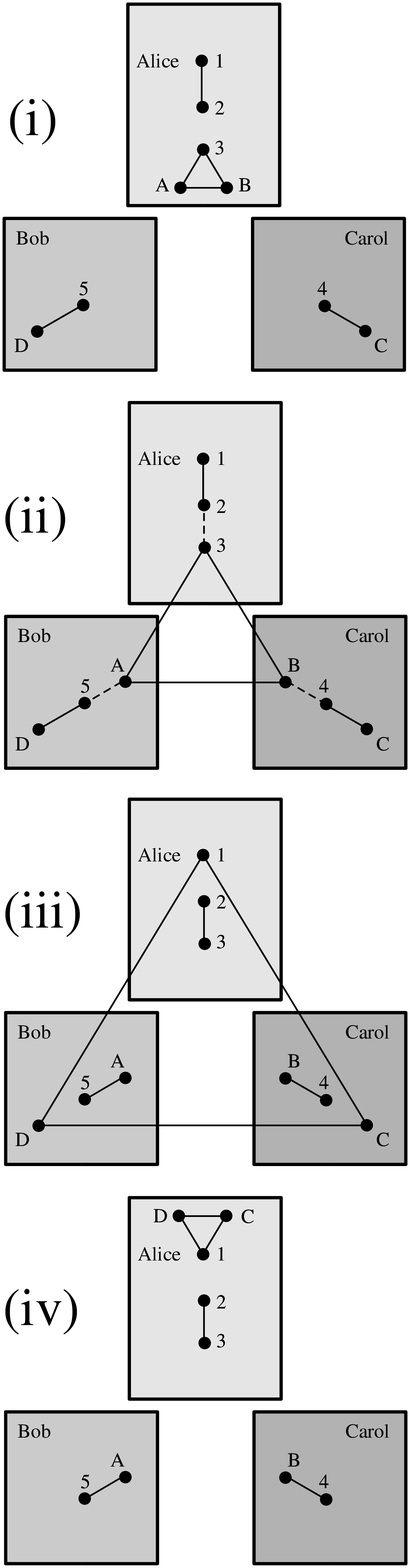}
\end{figure}
\noindent FIG..~2: {\small The four steps of the three-party
key distribution protocol based on ES.
The notation is the same introduced in Ref.~\cite{BVK98}:
triangles connect qubits in GHZ states,
bold lines connect qubits in Bell states, and
dashed lines represent Bell-state measurements.}


\begin{references}
\bibitem{com} In this protocol,
the results of the public measurement play
the same role as Alice's choice between bases
in the protocol of Ref.~\cite{BB84}.
Therefore, in addition to those mentioned
in Ref.~\cite{Cabello00b},
one interesting feature of the protocol based on ES is that
the ``choice'' of basis is truly random, and
therefore it does not require
any specific set-up to randomize it.
\bibitem{com2} Steps 1 and 2 of the particular
attack discussed in Ref.~\cite{Cabello00b} are equivalent to this
procedure.
\end{references}
\end{document}